\documentclass[twoside,twocolumn,english,prc,amsmath,a4paper,myheadings]{revtex4}
\usepackage[T1]{fontenc}
\usepackage{geometry}
\usepackage{amsmath}
\usepackage{graphicx}
\usepackage{amssymb}

\makeatletter
\def\@oddhead{\rightmark \hfill  Bose-Einstein Correlations in a Fluid Dynamical Scenario for Prototon-Proton Scattering at 7 TeV  \hfill \thepage}
\def\@evenhead{\thepage \hfill K. Werner et al.\hfill}
\def\fnum@table{\tablename~{\bf\thetable}}
\def\fnum@figure{\figurename~{\bf\thefigure}}
\def\tablename{\footnotesize{\bf Table}}
\def\figurename{\footnotesize{\bf Figure}}

\usepackage{dcolumn}

\def\citet{\cite}

\makeatother

\usepackage{babel}

\begin{document}

\title{{\normalsize Bose-Einstein Correlations in a Fluid Dynamical Scenario
for Proton-Proton Scattering at 7 TeV}}

\author{{\normalsize K.$\,$Werner$^{(a)}$,K. Mikhailov$^{(b)}$, Iu.$\,$Karpenko$^{(c)}$,
T.$\,$Pierog$^{(d)}$}}

\address{$^{(a)}$ SUBATECH, University of Nantes -- IN2P3/CNRS-- EMN, Nantes,
France}

\address{$^{(b)}$ Institute for Theoretical and Experimental Physics, Moscow,
117218, Russia}

\address{$^{(c)}$ Bogolyubov Institute for Theoretical Physics, Kiev 143,
03680, Ukraine}

\address{$^{(d)}$ Karlsruhe Institut of Technology (KIT), Institut fuer Kernphysik,
Karlsruhe, Germany}

\begin{abstract}
Using a fluid dynamical scenario for $pp$ scattering at 7 TeV, we
compute correlation functions for $\pi^{+}\pi^{+}$ pairs. Femtoscopic
radii are extracted based on three-dimensional parametrizations of
the correlation functions. We study the radii as a function of the
transverse momenta of the pairs, for different multiplicity classes,
corresponding to recent experimental results from ALICE. We find the
same decrease of the radii with $k_{T}$, more and more pronounced
with increasing multiplicity, but absent for the lowest multiplicities.
In the model we understand this as transition from string expansion
(low multiplicity) towards a three-dimensional hydrodynamical expansion
(high multiplicity). 
\end{abstract}
\maketitle

\section{Introduction}

Bose-Einstein correlations have proven to be a very useful tool to
provide space-time information about colliding systems at relativistic
energies. Different kinds of reactions have been considered, elementary
ones like electron-positron annihilation \citet{femtoee}, or more
complex systems like proton-proton and nucleus-nucleus collisions
\citet{femtoaa}. Sophisticated methods have been developed in particular
for heavy ion collisions at 200 GeV, to interpret the dependence of
the two particle correlation function $C$ on the relative pair momentum
$\mathbf{q}$. A simple way to summarize the results amounts to parametrize
the correlation function as\begin{align}
 & C(\mathbf{q})=\\
 & \quad1+\lambda\,\exp\left(-R_{\mathrm{out}}^{2}\, q_{\mathrm{out}}^{2}-R_{\mathrm{side}}^{2}\, q_{\mathrm{side}}^{2}-R_{\mathrm{long}}^{2}\, q_{\mathrm{long}}^{2}\right),\nonumber \end{align}
where \char`\"{}long\char`\"{} refers to the beam direction, \char`\"{}out\char`\"{}
is parallel to the projection of the pair momentum $\mathrm{\mathbf{P}}$
perpendicular to the beam, and \char`\"{}side\char`\"{} is the direction
orthogonal to \char`\"{}long\char`\"{} and \char`\"{}out\char`\"{}
\citet{fto-coord1,fto-coord2,fto-coord3}. The fit parameters $R_{\mathrm{out}}$,
$R_{\mathrm{side}}$, and $R_{\mathrm{long}}$ will be referred to
as {}``femtoscopic radii'' in the following. In this way, one can
study for example the dependence of these radii on the centrality
of the reaction, which is interesting because for central collisions
we expect higher energy densities and finally more collective flow
compared to peripheral collisions. A very general feature in heavy
ion collisions seems to be the fact that all the radii increase with
centrality. It is also found that for a given centrality all radii
decrease substantially with increasing transverse momentum $k_{T}=|\mathbf{P}_{T}/2|$.
This is compatible with a scenario of collective flow with a strong
space-momentum correlation, as we are going to discuss later.

Recently first results have been shown concerning the $k_{T}$ dependence
of the femtoscopic radii, for different multiplicity classes \citet{alice}.
The amazing result: with increasing multiplicity, one observes a more
and more visible decrease of the radii with $k_{T}$, as in heavy
ions. And the radii increase with multiplicity, similar to the increase
of the radii with centrality in heavy ion collisions. So do we have
the same dynamics which governs proton-proton and heavy ion collisions?
Is there a collective expansion driven by hydrodynamics in pp scattering
as well? 

To contribute to the answer to these questions, we will postulate
a scenario of hydrodynamic evolutions based on flux-tube initial conditions,
and compute correlation function, make the same fitting procedures
as in the experiment, and analyze the $k_{T}$ dependence of the femtoscopic
radii.

\section{Hydrodynamic scenario}

Originally hydrodynamics was only thought to present a valid description
for almost central collisions of heavy nuclei, where the volume is
(relatively) big. But it seems that this approach works very well
for all centralities. There is also no fundamental difference seen
between CuCu and AuAu, although the copper system is much smaller.
So it seems that systems much smaller than central AuAu fit well into
this fluid picture. Finally it is more and more accepted that the
famous ridge structure observed in angle-rapidity dihadron correlation
\citet{cms_ridge} is due to fluctuating initial conditions, which
are subsequently transformed into collective flow \citet{epos2}.
Here, the relevant scale for applying hydrodynamics is not the nuclear
size, but the size of the fluctuations, which is typically 1-2 fm.

Hydrodynamics is derived from a gradient expansion. In order to justify
a hydrodynamical treatment one has to relate the length scale with
the viscosity $\eta$. There are many estimates of numerical values
for the $\eta/S$ ($S$ being the entropy density), however, all are
based on unproven model assumptions. As shown in \citet{epos2}, the
variation of the ratio of elliptical flow to eccentricity can be perfectly
explained based on ideal hydrodynamics, whereas other authors extract
a non-vanishing viscosity from the same observable. It should also
be mentioned that in \citet{epos2}, the ideal hydrodynamical partonic
phase is followed by a highly viscous hadronic one, so in the average
the system is viscous. By talking about ideal hydrodynamics, we mean
an ideal hydrodynamical partonic phase, and as shown in \citet{epos2},
at least all RHIC AuAu data on soft physics are best described with
viscosity zero. From a theoretical point of view, the viscosity of
a QGP is unknown, and even a lower limit for $\eta/S$ as $1/4\pi$
is not a mathematical limit, but rather an estimate with unknown error
compared to the QCD value.

Is QGP formation a nuclear phenomenon? Or can it be formed in pp scattering,
as already proposed in \citet{oldhydro1,oldhydro2,oldhydro3,oldhydro4,kw-pp09,kw-ridge}?
Based on the above discussion, there is no reason not to treat proton-proton
scattering in the same way as heavy ions, namely incorporating a hydrodynamical
evolution. This approach makes clear predictions for many variables,
so the Nature will tell us whether the approach is justified or not.
Therefore it will be extremely interesting to think about the implications
of such a mini QGP, how such a small system can equilibrate so quickly,
and so on. It would be an enormous waste of opportunities, not to
consider this possibility, since a vast amount of proton-proton data
will be available very soon, concerning all kinds of observables.

What makes pp scattering at LHC energies interesting in this respect,
is the fact that at this high energy multiple scattering becomes very
important, where a large number of scatterings amounts to a large
multiplicity. In such cases, very large energy densities occur, even
bigger than the values obtained in heavy ion collisions at RHIC --
but in a smaller volume. Several authors discussed already the possibility
of a hydrodynamical phase in pp collisions at the LHC, to explain
the ridge correlation or to predict elliptical flow \citet{v2pp1,v2pp2,v2pp3,v2pp4,v2pp5,kw-ridge}.

We are going to employ a sophisticated hydrodynamical scenario, first
presented in ref. \citet{epos2} where many details can be found,
with the following main features: 

\begin{itemize}
\item initial conditions obtained from a flux tube approach (EPOS), compatible
with the string model used since many years for elementary collisions
(electron-positron, proton proton), and the color glass condensate
picture \citet{cgc1}; 
\item event-by-event procedure, taking into the account the highly irregular
space structure of single events, being experimentally visible via
so-called ridge structures in two-particle correlations; 
\item core-corona separation, considering the fact that only a part of the
matter thermalizes \citet{kw-core}; only in the core region, the
energy density from the strings is considered for the hydrodynamical
evolution;
\item use of an efficient code for solving the hydrodynamic equations in
3+1 dimensions, including the conservation of baryon number, strangeness,
and electric charge; 
\item employment of a realistic equation-of-state, compatible with lattice
gauge results -- with a cross-over transition from the hadronic to
the plasma phase \citet{lattice,lattice-k}; 
\item use of a complete hadron resonance table, making our calculations
compatible with the results from statistical models;
\item hadronic cascade procedure after hadronization from the thermal system
at an early stage \citet{urqmd,urqmd2}.
\end{itemize}
In ref. \citet{epos2}~$\!$, we test the approach by investigating
all soft observables of heavy ion physics, in case of AuAu scattering
at 200 GeV. In refs. \citet{kw-pp09,kw-ridge} we investigate first
proton-proton results, with among other things {}``the ridge''.

\begin{figure}[tb]
\begin{centering}
\hspace*{-0.4cm}\includegraphics[scale=0.27]{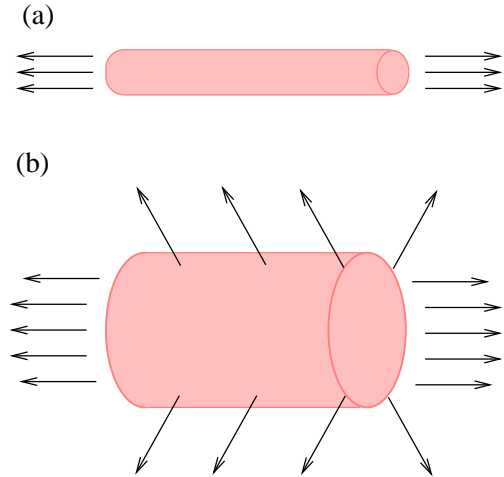}
\par\end{centering}

\begin{centering}
\caption{(Color online) Expansion of a low multiplicity (a) and a high multiplicity
(b) event.\label{cap:flux}}

\par\end{centering}
\end{figure}

\section{Results}

\begin{figure}[tb]
\begin{centering}
\includegraphics[angle=270,scale=0.35]{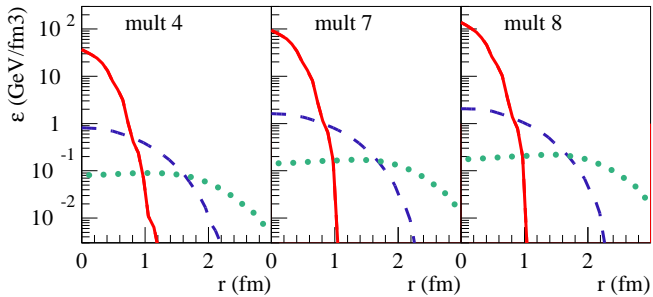}\caption{(Color online) Evolution of the energy density (average over many
events) for the three multiplicity classes \textsf{\emph{Milt~4}},
\textsf{\emph{Milt~7}} and \textsf{\emph{Milt~8}}. The full red
lines are the initial conditions, the other curves are respectively
1 and 2 fm/c later.\label{cap:eps}}

\par\end{centering}
\end{figure}

In the following, we will consider several multiplicity classes, named
\textsf{\emph{mult~1}}, \textsf{\emph{mult~4}} , \textsf{\emph{mult~7}}
and \textsf{\emph{mult~8}}, corresponding to four out of the eight
multiplicity classes used in ref. \citet{alice}, going from low multiplicity
(\textsf{\emph{mult~1}}, less than minimum bias) to high multiplicity
(\textsf{\emph{mult}}~8, five time minimum bias). Our core-corona
procedure will find no core for \textsf{\emph{mult~1,}}\textsf{ }then
increasing core fraction, and for\textsf{ }\textsf{\emph{mult~4}}
to \textsf{\emph{mult~8 }}\textsf{e}ssentially core, with increasing
energy densities. So the\textsf{\emph{ mult~1 }}events are just ordinary
strings which expand longitudinally (see fig. \ref{cap:flux}(a)),
whereas\textsf{\emph{ mult~4}} to \textsf{\emph{mult~8 }}show a
hydrodynamical expansion, also in transverse direction (see fig. \ref{cap:flux}(b)).
In fig. \ref{cap:eps}, we show the evolution of the energy density
for the different multiplicity classes. Obviously the energy density
increases with multiplicity, values of more than 100 GeV/fm$^{3}$
are achieved. %
\begin{figure}[b]
\begin{centering}
\includegraphics[angle=270,scale=0.35]{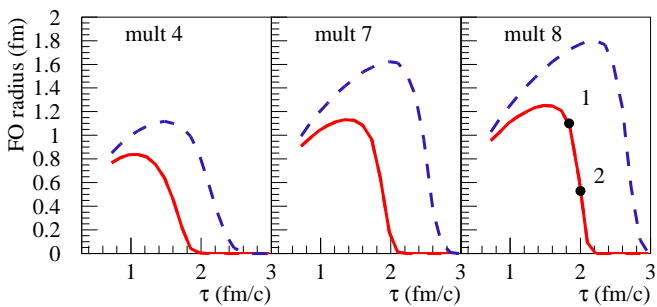}\caption{(Color online) Freeze out radii, using the EoS from \citet{lattice}
(solid curves), and using \citet{lattice-k} (dashed curves).\label{cap:fo}}

\par\end{centering}
\end{figure}
We consider two options for the equation of state, one being a parametrization
of the results of \citet{lattice}, the other one referring to \citet{lattice-k}.
There is a big difference between the two, the transition temperature
is much lower and the transition is much smoother in\citet{lattice},
compared to \citet{lattice-k}. The corresponding {}``freeze out''
radii (where the transition hydro / cascade is done) are very different,
see fig. \ref{cap:fo}. The final results will differ less, because
the early freeze out for the EoS from \citet{lattice} is followed
by an intense hadronic rescattering.

\begin{figure}[b]
\begin{centering}
\includegraphics[scale=0.29]{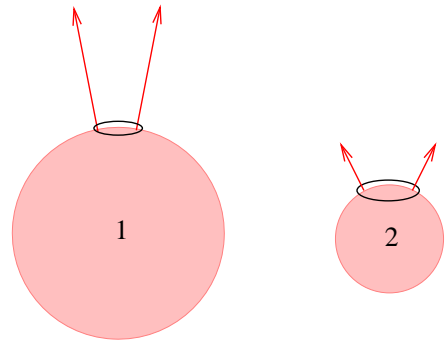}\caption{(Color online) Emitting pairs with fixed momentum difference in radial
direction, from different radii.\label{cap:fto}}

\par\end{centering}
\end{figure}

Based on roughly ten million simulations of the hydrodynamical evolution,
we compute in the usual way the correlation functions for $\pi^{+}\pi^{+}$
pairs, taking into account Bose-Einstein statistics, as discussed
in \citet{epos2,kw-pp09,hbt-lednicki}. Whereas in the data Pythias
has to be used as {}``baseline'', we stay consistently within our
scenario and use simply a calculation without Bose-Einstein statistics
as baseline. We then fit the correlation functions to obtain the radii.
Before showing the results, let us discuss in a qualitative way why
we expect a decrease of the radii with $k_{T}$. Let us consider the
freeze out curves of fig. \ref{cap:fo}: Most particle production
occurs in the region where the radii drop to zero. Comparing the two
points {}``1'' and {}``2'' in the figure, we have to recall that
the collective flow at a large radius (1) is much bigger compared
to small radius (2). In fig. \ref{cap:fto}, we sketch the corresponding
momentum vectors of pairs, emitted at large and small radii (which
finally amounts to large and small $k_{T}$), where the vectors are
such that their difference is the same. The space distances in case
2 are then bigger than the ones in case 1 (and these distances are
essentially the femtoscopic radii). 

\begin{figure*}[tb]
\begin{centering}
\includegraphics[angle=270,scale=0.5]{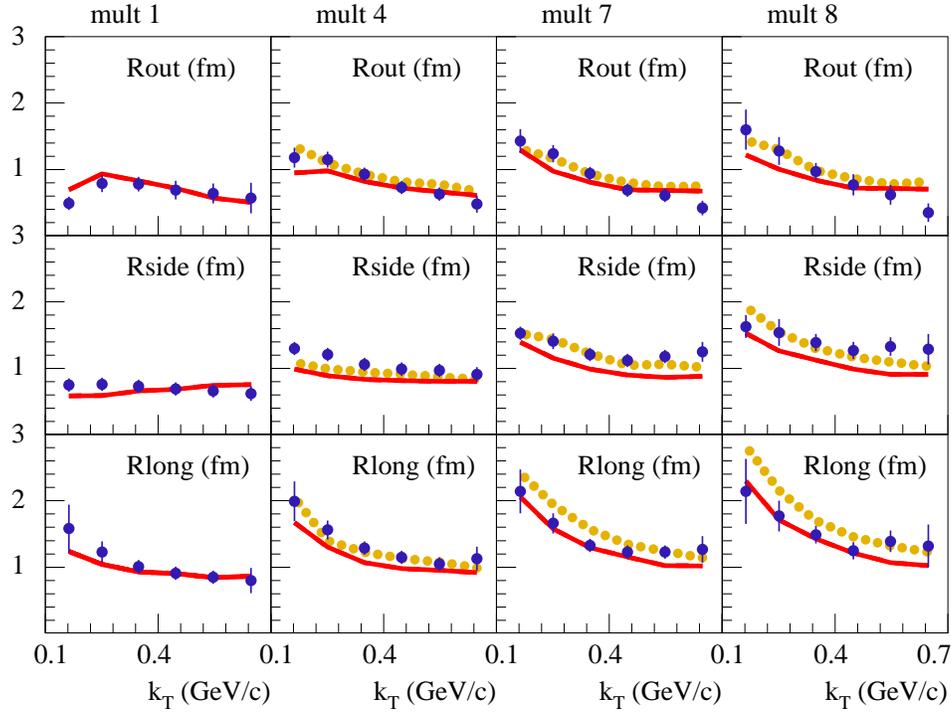}\caption{(Color online) Femtoscopic radii for three different multiplicity
classes, using an EoS compatible with ref. \citet{lattice} (solid
curves) or compatible with ref. \citet{lattice-k} (dotted curves).
The curves are the same for \textsf{\emph{Milt~1}} because there
is no fluid phase. The points are data. \label{cap:final} The curves
are absolute predictions, the parameters of the model are obtained
from other comparisons (yields, $p_{t}$ spectra). }

\par\end{centering}
\end{figure*}

Let us discuss our main result: As seen in fig. \ref{cap:final},
all radii indeed show a more and more pronounced decrease with increasing
$k_{T}$, for data and simulations, which can -- in the calculations
-- clearly be attributed to collective flow. For the case \textsf{\emph{Milt~1}}
the radii $R_{\mathrm{out}}$ and $R_{\mathrm{side}}$ are essentially
flat, only $R_{\mathrm{long}}$ has already some $k_{T}$ dependence.
So we see here nicely the transition from a longitudinal expansion
(string) towards a three-dimensional hydrodynamical expansion for
higher multiplicities. %

\begin{acknowledgments}
This research has been carried out within the scope of the ERG (GDRE)
{}``Heavy ions at ultra-relativistic energies'', a European Research
Group comprising IN2P3/CNRS, Ecole des Mines de Nantes, Universite
de Nantes, Warsaw University of Technology, JINR Dubna, ITEP Moscow,
and Bogolyubov Institute for Theoretical Physics NAS of Ukraine. Iu.K.
acknowledges partial support by the State Fund for Fundamental Researches
of Ukraine (Agreement of 2011) and National Academy of Sciences of
Ukraine (Agreement of 2011). T.P. and K.W. acknowledge partial support
by a PICS (CNRS) with KIT (Karlsruhe). K.M. acknowledges partial support
by the RFBR-CNRS grants No 10-02-93118-NTsNIL\_a and 10-02-93111-NTsNIL\_a.
\end{acknowledgments}

\end{document}